\documentclass{article}
\usepackage{amsfonts,amsmath}
\newcommand{\be}{\begin{equation}}
\newcommand{\ee}{\end{equation}}
\newcommand{\bs}{\begin{split}}
\newcommand{\es}{\end{split}}

\newcommand{\bea}{\begin{eqnarray}}
\newcommand{\ol}{\overline}
\newcommand{\eea}{\end{eqnarray}}

\newcommand{\p}{\partial}

\newcommand{\m}{\mu}

\newcommand{\nc}{\Theta}

\newcommand{\vf}{\varphi}

\newcommand{\hgamma}{\hat{\gamma}}

\newcommand{\hb}{\hat{b}}

\newcommand{\diff}{\mathrm{d}}

\newcommand{\rmS}{\mathrm{S}}
\newcommand{\AdS}{$\mathrm{AdS}_5 \times \mathrm{S}^5$}
\newcommand{\Ncal}{\mathcal{N}}
\newcommand{\Gcal}{\mathcal{G}}
\newcommand{\Ocal}{\mathcal{O}}
\newcommand{\Wcal}{\mathcal{W}}
\newcommand{\Tr}{\mathrm{Tr}}
\newcommand{\Tcal}{\mathcal{T}}
\newcommand{\Vcal}{\mathcal{V}}

\DeclareSymbolFont{AMSa}{U}{msa}{m}{n}
\DeclareSymbolFont{AMSb}{U}{msb}{m}{n}
\DeclareMathSymbol{\fieldR}{\mathalpha}{AMSb}{"52}
\newif\ifpdf
\ifx\pdfoutput\uundefined
\pdffalse 
\else
\pdfoutput=1 
\pdftrue \fi

\begin{document}
\begin{titlepage}
\begin{center}
\hfill {\tt YITP-SB-06-45}\\
\vskip 20mm

{\Large {\bf On $\beta$--deformations and Noncommutativity}}

\vskip 10mm

{\bf Manuela Kulaxizi}

\vskip 4mm {\em C. N. Yang Institute for Theoretical
Physics}\\
{\em State University of New York at Stony Brook}\\
{\em Stony Brook, NY 11794-3840, USA}\\
[2mm] E-mail: {\tt kulaxizi@insti.physics.sunysb.edu}
\end{center}
\vskip 1in

\begin{center} {\bf ABSTRACT }\end{center}

\vskip 4pt  We elucidate the connection between the $\Ncal=1$ $\beta$--deformed SYM theory and noncommutativity.
Our starting point is the T--duality generating transformation involved in constructing the gravity duals of both 
$\beta$--deformed and noncommutative gauge theories. We show that the two methods can be identified provided 
that a particular submatrix of the $O(3,3,\mathbb{R})$ group element employed in the former case, is interpreted 
as the noncommutativity parameter associated with the deformation of the transverse space. It is then explained
how to construct the matrix in question, relying solely on information extracted from the gauge 
theory Lagrangian and basic notions of AdS/CFT. This result may provide an additional tool in exploring deformations 
of the $\Ncal=4$ SYM theory. Finally we use the uncovered relationship between $\beta$--deformations and noncommutativity 
to find the gravity background dual to a noncommutative gauge theory with $\beta$--type noncommutativity parameter.

\begin{quotation}
\noindent

\end{quotation}
\vfill \flushleft{October 2006}
\end{titlepage}
\eject

\section{Introduction}

The conjectured gauge/gravity duality \cite{Maldacena}\cite{Gubseretal98}\cite{Witten9802} relates four--dimensional theories at 
strong t'Hooft coupling with weakly coupled gravitational ones. 
In \cite{LuninMaldacena} Lunin and Maldacena presented a further development in this direction
by constructing the gravity duals of gauge theories deformed in a particular manner that maintains 
a global $U(1)\times U(1)$ symmetry present in the original undeformed theory. The prototype of these 
deformations is a Leigh--Strassler \cite{LeighStrassler95} exactly marginal deformation of $\Ncal=4$ 
SYM theory, characterized by a complex parameter $\beta$ which preserves $\Ncal=1$ 
supersymmetry. The method of Lunin and Maldacena is not however restricted to 
conformal field theories. It can be applied to any field theory as long as its dual
gravity background contains a two torus geometrically realizing the global $U(1)$ 
symmetries in question. When $\beta\in\mathbb{R}$ --- usually denoted as $\gamma$ in 
the literature --- the prescription presented in \cite{LuninMaldacena} amounts to performing 
an $SL(2,\mathbb{R})$ transformation on the complexified K\"ahler modulus $\tau$ of this two torus. 
The specific element of  $SL(2,\mathbb{R})$ under consideration has only one free parameter 
which is then identified with the real deformation parameter $\gamma$ of the gauge theory. 
Subsequent work on the subject of the $\beta$--deformed gauge theories has provided
further checks of the AdS/CFT correspondence \cite{Frolovetal0503}\cite{Frolovetal0507}\cite{ChenKumar}
\cite{Durnfordetal}\cite{GeorgiouKhoze} \cite{Hernandezetal} whereas the possibility of an underlying integrable
structure in this context was explored in \cite{BerensteinCherkis04}\cite{BeisertRoiban}\cite{Frolov}. 
Several aspects of these deformations were analysed from the gauge theory viewpoint in 
\cite{Khoze}\cite{FreedmanGursoy}\cite{KuzenkoTseytlin}\cite{Rossietal05}\cite{Elmettietal}\cite{Rossietal06}\cite{Maurietal}. 
Furthermore, generalizations as well as applications of the solution generating technique introduced 
in \cite{LuninMaldacena} were considered in \cite{GursoyNunez}\cite{Frolov}\cite{Hernandezetal}
\cite{AhnVazquezPoritz05}\cite{AhnVazquezPoritz06}\cite{Rashkovetal}. 

Meanwhile, it became clear \cite{Aybike} that embedding $SL(2,\mathbb{R})$ into the T--duality group 
$O(2,2,\mathbb{R})$ may be a significantly easier way to obtain the deformed backgrounds since it 
suffices then to consider the action of the appropriate $O(2,2,\mathbb{R})$ group element on the 
background matrix $E=g+B$.
In this framework, an extraordinary similarity between the proposal of \cite{LuninMaldacena}
and the method for constructing gravity duals of noncommutative gauge theories
becomes evident \footnote{Actually, this connection was already noted in \cite{LuninMaldacena}.}.
From the gauge theory point of view this analogy is not surprising since the deformation
amounts to modifying the commutator of the matter fields in the Lagrangian or
equivalently, their product. A natural proposal for the product rule  
was set forth in \cite{LuninMaldacena} and subsequently verified in the
dual field theory context in \cite{Khoze}\cite{Durnfordetal}. 

The central aim of this note is to clarify the relation
between noncommutativity and $\beta$--deformations. We will consider the 
deformations in their original context as marginal deformations of $\Ncal=4$ SYM
and show how to obtain a noncommutativity matrix $\nc$ describing them.
The main point will be to think of the matter fields in the dual
theory as coordinates parametrizing the space transverse to the D3--brane
where the gauge theory lives. Then, reality properties, global symmetries and 
marginality will severely constrain the form of the noncommutativity 
matrix leaving one possible choice, the one which leads to the correct
gravity dual description. In other words, $\nc^{ij}$ along with the metric of the 
transverse space can be thought of as another way to encode the 
moduli space of the gauge theory. 
This suggests an alternative way in which to investigate deformations of 
the original AdS/CFT proposal \cite{Maldacena} by determining the
\emph{open} string parameters pertaining to them. Related ideas
will be explored in a forthcoming publication \cite{Kulaxizi0611} in order to study 
another Leigh--Strassler marginal deformation of $\Ncal=4$ SYM the gravity dual
of which is yet unkown.

The plan of this paper is as follows. 
In the next section, we review the solution generating technique proposed in \cite{LuninMaldacena}
as well as its formulation through T--duality \cite{Aybike}. 
In section \ref{NC}, we present some basic facts about noncommutative geometry. Then we describe the methods 
employed in finding the gravity duals of these theories in a fashion that makes evident the 
similarity with the approach of \cite{LuninMaldacena}. In particular, it is shown that the T--duality 
group elements used in both cases can be identified if the deformation submatrix referred to as $\mathbf{\Gamma}$ 
in \cite{Aybike} is interpreted as a noncommutativity matrix. 
In section \ref{BetaNC}, we explain how one can determine a suitable noncommutativity matrix for the 
$\beta$--deformed gauge theory. This construction is purely based on gauge theory data and basic notions of AdS/CFT. 
We then show that $\nc^{ij}$ is precisely the submatrix $\mathbf{\Gamma}$ appearing in section 1.
As an obvious way to exploit the precise relation uncovered between noncommutativity and $\beta$--deformations, 
we proceed to construct the gravity dual of a noncommutative gauge theory with $\beta$--type noncommutativity 
both in Euclidean and in Lorentzian signature. We finally present our conclusions in section \ref{Conclusion}.

\section{The Lunin--Maldacena solution generating technique.}\label{LM}

As it was shown in \cite{LeighStrassler95} $\Ncal=4$ Super Yang Mills
admits a complex three parameter family of marginal deformations preserving
$\Ncal=1$ supersymmetry which is described by the following superpotential:
\be\label{Superpotential}
\Wcal=\kappa \epsilon_{IJK}\Tr\left([\Phi^{I},\Phi^{J}]_{\beta}\Phi^{K}\right)+
\rho \Tr\left(\sum_{I=1}^{3}(\Phi^{I})^{3}\right)
\ee
Here $\Phi^{I}$ are three chiral superfields and  
$[\Phi^{I},\Phi^{J}]_{\beta}\equiv e^{i \beta} \Phi^{I} \Phi^{J}-e^{-i \beta} \Phi^{J} \Phi^{I}$.
Together with the gauge coupling $g_{YM}$, the complex parameters ($\kappa, \beta, \rho$) 
constitute the four couplings of the theory. Conformal invariance imposes one condition
on these couplings thus (\ref{Superpotential}) describes a three
parameter family of deformations. 
When $\rho=0$ the theory is often referred to as the $\beta$--deformed
gauge theory and preserves an additional global $U(1)\times U(1)$ symmetry 
(apart from the $U(1)_{R}$ R--symmetry) which acts on the superfields as 
follows:
\be\label{U1}\begin{split}
U(1)_{1}&:\quad \left(\Phi_{1},\Phi_{2},\Phi_{3}\right)\rightarrow(\Phi_{1},e^{i \alpha_{1}}\Phi_{2},e^{-i\alpha_{1}}\Phi_{3})\\
U(1)_{2}&:\quad \left(\Phi_{1},\Phi_{2},\Phi_{3}\right)\rightarrow(e^{-i \alpha_{2}}\Phi_{1},e^{i \alpha_{2}}\Phi_{2},\Phi_{3})
\end{split}
\ee
In this paper we will be mainly considering the $\beta$--deformed theory
for $\beta\in\mathbb{R}$.
It is then customary to denote the deformation parameter as $\gamma$ and we will
adhere to this notation in this section.
Lunin and Maldacena in \cite{LuninMaldacena} succeeded in finding the
gravity dual of this theory 
by implemeting a generating solution technique which can be
applied to any field theory with $U(1)\times U(1)$ global
symmetry realized geometrically.
Their method essentially consists in performing an $SL(2,\mathbb{R})$
transformation on the complexified K\"ahler modulus of the two torus
associated with the U(1) symmetries in question.
Suppose for instance that one knows the gravity dual of the
undeformed theory and furthermore that the two global U(1)'s 
of the parent theory also preserved by the deformation are indeed 
realized geometrically. Then the supergravity dual of the deformed 
theory is given by the following substitution:
\be\label{kahlermodulus}
\tau=(B_{12}+\sqrt{g})\rightarrow\frac{\tau}{1+\gamma\tau}
\ee
where $\tau$ is the complexified K\"ahler modulus of the two torus (associated to
the U(1) symmetries of the original solution)with $B_{12}$ the B--field along the torus
and $\sqrt{g}$ its volume. In other words, one considers the theory compactified on the two 
torus and subsequently acts on its K\"ahler modulus with the particular 
element of $SL(2,\mathbb{R})$ given by
$\left(\begin{smallmatrix}
a&b\\
c&d\\
\end{smallmatrix}\right)\equiv
\left(\begin{smallmatrix}
1&0\\ 
\gamma&1\\
\end{smallmatrix}\right)$ 
with $\gamma$ the parameter of the theory. This element of $SL(2,\mathbb{R})$ 
is chosen because it ensures that the new solution will present no singularities 
as long as the original metric is non--singular. An alternative way of thinking
about this solution generating transformation is in terms of applying
a series of T--dualities. More precisely, the method illustrated above is equivalent 
to doing a T--duality on a circle, a coordinate transformation and then another T--duality (TsT). 

Subsequently it was shown \cite{Aybike} that one can embed the $SL(2,\mathbb{R})$ 
that acts on the K\"ahler modulus into the T--duality group $O(2,2,\mathbb{R})$ 
and thus consider the action of the latter on the background matrix $E=g+B$. 
This provides a considerably simpler way of obtaining the new solutions.
For $\left(\begin{smallmatrix}
a&b\\
c&d\\
\end{smallmatrix}\right)$ the generic element of $SL(2,\mathbb{R})$ the appropriate 
embedding is the following:
\be\label{Telement}
\Tcal=\begin{pmatrix}
\mathbf{A}&\mathbf{B}\\
\mathbf{C}&\mathbf{D}\\
\end{pmatrix}=\begin{pmatrix}
a&0&0&b\\
0&a&-b&0\\
0&-c&d&0\\
c&0&0&d\\
\end{pmatrix}
\ee
It is then easy to see \cite{GiveonPorratiRabinovici} that $\Tcal$ transforms the 
original background matrix $E_{0}$ as:
\be
E_{0}\rightarrow E=(\mathbf{A}E_{0}+\mathbf{B})(\mathbf{C}E_{0}+\mathbf{D})^{-1}\equiv\frac{\mathbf{A}E_{0}+\mathbf{B}}{\mathbf{C}E_{0}+\mathbf{D}}
\ee
where the $2\times 2$ matrices A,B,C,D are defined through (\ref{Telement}).
According to \cite{LuninMaldacena} we should not consider any $SL(2,\mathbb{R})$ 
element but the precise one with $a=d=1$, $b=0$ and $c=\gamma$. In this case (\ref{Telement}) 
reads:
\be\label{O22} 
\Tcal=
\begin{pmatrix}
\mathbf{1}& \mathbf{0}\\
\mathbf{\Gamma}& \mathbf{1}\\
\end{pmatrix}\quad \text{with}\quad
\mathbf{\Gamma}=
\begin{pmatrix}
0  & -\gamma\\
\gamma & 0 \\
\end{pmatrix}
\ee
where $\mathbf{1}$  and  $\mathbf{0}$ represent the $2 \times 2$ identity and zero 
matrices respectively.
Following now the T--duality rules in \cite{GiveonPorratiRabinovici} 
we can write the NS--NS fields of the new solution in terms of $E_{0}$ and $\Gamma$ 
as follows:
\be\label{NSFieldsTransform}
\begin{split}
E&=\frac{1}{E_{0}^{-1}+\mathbf{\Gamma}}\\
e^{2\Phi}&=\det({1+E_{0}\mathbf{\Gamma}})e^{2\Phi_{0}} 
\end{split}\ee
The RR-fields of the background can be obtained in a similar fashion
using the transformation rules of 
\cite{Sundell}\cite{Bergshoeffetal}\cite{Cveticetal}\cite{Fukumaetal}\cite{Hassan}. 
Nevertheless, for the purposes of this letter it only suffices to know that appropriate rules 
exist and can be applied. 

There are however cases where one needs to slightly modify the method
illustrated above. This happens when non--trivial fibrations 
mix the isometry diretions of the two torus with other directions in the metric.
It is then necessary to embed $SL(2,\mathbb{R})$ into $O(n+2,n+2,\mathbb{R})$ 
with n the number of non--trivial coordinate fibrations. A particular example of 
this is the $\mathrm{AdS}_{5}\times\mathrm{T}^{1,1}$ solution of \cite{KlebanovWitten}.
If we want to apply the deformation to this background instead of (\ref{O22})
we should employ:
\be\label{O33} 
\Tcal=\begin{pmatrix}
\mathbf{1}& \mathbf{0}\\
\mathbf{\Gamma}& \mathbf{1}\\
\end{pmatrix}\quad \text{where} \quad 
\mathbf{\Gamma}=\begin{pmatrix}
0 & -\gamma & 0\\
\gamma & 0 & 0\\
0 & 0 & 0\\
\end{pmatrix}
\ee
Furthermore, as it was again pointed out in \cite{Aybike}, the appropriate 
T--duality matrix one should use for the deformation of \AdS    which gives
rise to the gravity dual of the $\beta$--deformed gauge theory is:
\be\label{O33LM} 
\Tcal=\begin{pmatrix}
\mathbf{1}& \mathbf{0}\\
\mathbf{\Gamma}& \mathbf{1}\\
\end{pmatrix}\quad \text{where now}\quad 
\mathbf{\Gamma} =\begin{pmatrix}
0 & -\gamma & \gamma\\
\gamma & 0 & -\gamma\\
-\gamma & \gamma & 0\\
\end{pmatrix}
\ee
This particular choice of $\mathbf{\Gamma}$ with the necessary embedding of $SL(2,\mathbb{R})$ 
into $O(3,3,\mathbb{R})$ can be understood in this case as the result of performing a 
change of coordinates and a T--duality transformation of the form (\ref{O33}) followed
by another coordinate transformation \cite{Aybike}.
For future reference and as a concrete illustration of the above we would like to 
give an explicit construction of the background in this case.
What we have to do is to simply act with (\ref{O33LM}) on the background matrix $E_{0}$ which 
in this example is none other but \AdS. Since we are interested in obtaining the gravity dual 
of a conformal gauge theory we expect that only the $S^{5}$ part of \AdS  will be 
affected by the deformation. We can write the metric on $S^{5}$ in the following way: 
\be
ds^2=R^2 \left(\sum_{i=1}^{3}d\mu_{i}^2+\mu_{i}^2 d\phi_{i}^2\right)\quad \text{where}\quad\sum_{i=1}^3\mu_{i}^2=1
\ee
Note here that we want to deform the geometry along the U(1) isometry directions of $S^{5}$,
therefore the relevant part of the backgound matrix is:
\be
E_{0}=R^2 \begin{pmatrix}
\mu_{1}^2&0&0\\
0&\mu_{2}^2&0\\
0&0&\mu_{3}^2\\
\end{pmatrix}
\ee
Using now equation (\ref{NSFieldsTransform}) and its generalization fo RR--fields we find \cite{Aybike}:
\be \begin{split}\label{LMsolution} 
\diff s^2=&R^2(\diff s^2_{\mathrm{AdS}_{5}}+\diff s^2_{5}),\quad \mathrm{where:}\quad
\diff s^2_{5}=\sum_{i}(\diff\m^2_{i}+G
\m^2_{i}\diff\phi^2_{i})+\hgamma G
\m^2_{1}\m^2_{2}\m^2_{3}(\sum_{i}\diff\phi_{i})^2\\
G^{-1}&=1+\hgamma^2(\sum_{i\neq j}\m^2_{i}\m^2_{j}), \quad\quad \hgamma=\mathrm{R}^2\gamma, \quad\quad
\mathrm{R}^4=4 \pi e^{\Phi_{0}}\mathrm{N} \\
e^{2 \vf}&=e^{2 \vf_{0}}G, \quad B=\hgamma R^2 G \left(\sum_{i\neq j}\m^2_{i}\m^2_{j}\diff\phi_{i}\diff\phi_{j}\right)\\
C_{2}&=-\gamma(16 \pi N)\omega_{1}(\sum_{i}\diff\phi_{i}),\quad
C_{4}=(16 \pi N)(\omega_{4}+G \omega_{1}\diff\phi_{1}\diff\phi_{2}\diff\phi_{3})\\
F_{5}&=(16 \pi N)(\omega_{\mathrm{AdS}_{5}}+G \omega_{\rmS^{5}}),\quad
\omega_{\rmS^{5}}=\diff\omega_{1}\diff\phi_{1}\diff\phi_{2}\diff\phi_{3}, \quad 
\omega_{\mathrm{AdS}_{5}}=\diff\omega_{4}
\end{split}\ee
which is precisely the gravity solution given in \cite{LuninMaldacena}.

\section{The gravity duals of noncommutative gauge theories.}\label{NC}

In this section we would like to focus on yet another class of 
supergravity duals which can be obtained in manner analogous to
the one described earlier. These are the gravity duals of 
noncommutative gauge theories \footnote{For an introduction to 
noncommutative geometry see for example \cite{DouglasNekrasov} 
and references therein} and in fact the methodology used in both cases 
is almost identical.

Noncommutative --- as opposed to ordinary --- gauge theories,
live in a space of noncommuting coordinates \footnote{We limit the 
discussion in this section to Euclidean spaces or to 
noncommutativity which does not affect the time--like coordinate.}. 
Such a deformation of space is encoded in what is referred to as
the noncommutativity parameter $\nc^{ij}$ defined as:
\be\label{commutator}
[x^{i},x^{j}]=i \nc^{ij}
\ee 
where $\{x^{i}\}$ is a set of coordinates parametrizing 
the space and $\nc^{ij}$ a real antisymmetric matrix.
In general, the easiest way to deal with functions
on these spaces is to replace noncommuting variables 
with commuting ones by simply defining a new product
rule between them, usually called a star product. 
The star product will then contain all the information 
on the noncommutative structure of the space.

Out of all the possible forms of $\nc^{ij}$ the case 
most well understood is by far the one in which the
commutators of (\ref{commutator}) are c--numbers 
and therefore the noncommutativity parameter is 
essentially a constant. In this case, associativity is 
preserved and the appropriate star product has the form:
\be\label{Moyal}
f(x)\ast g(x)=f(x+\xi) e^{\frac{i}{2} \overleftarrow{\frac{\p}{\p\xi^{i}}} \nc^{ij} \overrightarrow{\frac{\p}{\p\zeta^{j}}}}g(x+\zeta)=f \left(1+\overleftarrow\p_{i}\nc^{ij}\overrightarrow\p_{j}+\Ocal(\nc^2)\right) g
\ee
Gravity duals of theories living on noncommutative spaces
with constant noncommutativity parameter were first 
found in \cite{MaldacenaRusso}\cite{HashimotoItzhaki}.
The basic technique for constructing these solutions 
is to combine diagonal T--dualities, constant shifts of 
the NS--NS two form and $SO(p,1)$ transformations, where
p is the number of spatial dimensions. One first T--dualizes in 
the directions where one wants to turn on fluxes, shifts the B 
field by a constant in these directions and then T--dualizes back. 
Equivalently, one can T--dualize along one of the directions 
of the fluxes, use a boost/rotation between a non compact and a 
compact direction and the T--dualize back. Both methods give the 
same result. 
It was later on realized that \cite{Bermanetal} these solutions
can be generated from the action of the $O(p,p,\mathbb{R})$ 
T--duality group element
\be
\Tcal=\begin{pmatrix}
\mathbf{1}&\mathbf{0}\\
\mathbf{\nc}&\mathbf{1}\\
\end{pmatrix}
\ee    
on the original undeformed solution where now $\mathbf{0},\mathbf{1},\mathbf{\nc}$ are $p$
dimensional square matrices with $p$ denoting the number of spatial directions along which 
noncommutativity is turned on. 
Suppose for instance that one wants to describe a gauge theory
living in four dimensional Euclidean space employed with cartesian 
coordinates $x^{\mu}$ where: 
$[x^{0},x^{1}]=i b_{1}$ and $[x^{2},x^{3}]=i b_{2}$. 
It is then clear that one should consider the embedding of the noncommutativity  
parameter into the T--duality group $O(4,4,\mathbb{R})$ as follows:
\be\label{TcalNC}
\Tcal=
\begin{pmatrix}
\mathbf{1}_{4}& \mathbf{0}_{4}\\
\mathbf{\nc}& \mathbf{1}_{4}\\
\end{pmatrix}\quad \text{with}\quad
\mathbf{\nc}=
\begin{pmatrix}
0&b_{1}&0&0\\
-b_{1}&0&0&0\\
0&0&0&b_{2}\\
0&0&-b_{2}&0\\
\end{pmatrix}
\ee
The original solution to be deformed in this context is again \AdS, however now $\nc$
lies along the non--compact, $\mathrm{AdS}_{5}$ piece of the geometry. Writting the 
metric on $\mathrm{AdS}_{5}$ as:
\be
ds_{AdS}^2=R^2 u^2 (dx_{0}^2+dx_{1}^2+dx_{2}^2+dx_{3}^2)+R^2\frac{du^2}{u^2}
\ee
we see that the relevant part of the background matrix $E_{0}$ in this case is:
\be
E_{0}=R^2 u^2\begin{pmatrix}
1&0&0&0\\
0&1&0&0\\
0&0&1&0\\
0&0&0&1\\
\end{pmatrix}
\ee
and acting now on $E_{0}$ with the T--duality matrix $\Tcal$ of equation (\ref{TcalNC})
we obtain \cite{MaldacenaRusso}:
\be\label{MaldacenaRusso}
\begin{split}
ds^2_{str}&=u^2 R^2(G_{1}(dx_{0}^2+dx_{1}^2)+ G_{2} (dx_{2}^2+dx_{3}^2))+\frac{R^2}{u^2}(du^2+u^2 d\Omega_{5}^2)\\
B&=\hb_{1} R^2 G_{1} u^4 dx_{0}\wedge dx_{1}+\hb_{2} R^2 G_{2} u^4 dx_{2}\wedge dx_{3} \\
e^{2\Phi}&=G_{1} G_{2} e^{2 \Phi_{0}} ,\quad G_{1}=\frac{1}{1+\hb_{1}^2 u^4}, \quad G_{2}=\frac{1}{1+\hb_{2}^2 u^4}\\
\hb_{1}&=R^2 b_{1},\quad \hb_{2}=R^2 b_{2} \end{split}
\ee
which is the gravity dual \footnote{Note the resemblance between (\ref{LMsolution}) and (\ref{MaldacenaRusso}).}
of a noncommutative gauge theory defined in Euclidean space with $[x^{0},x^{1}]=i b_{1}$ and $[x^{2},x^{3}]=i b_{2}$.
The Langrangian description of this theory can be easily derived from the
$\Ncal=4$ SYM Langrangian by replacing the ordinary product of 
functions with the Moyal star designated in (\ref{Moyal}).

Although we have so far considered applying this method directly to 
the near horizon geometry one can, perhaps even more appropriately, 
perform it on the p--brane solutions as well
\cite{Itzhakietal}\cite{Bermanetal}\cite{Alishahihaetal99}\cite{Alishahihaetal05}. 
The near horizon limit that needs to taken in this case requires a relative 
scaling between the B--field and the metric g which actually corresponds to 
the Seiberg--Witten limit proposed in \cite{SeibergWitten9908}.  

It should now be evident that the solution generating transform 
employed by Lunin and Maldacena in order to find the gravity duals of 
$\beta$--deformed gauge theories is almost identical to the one used
for the same purpose within the context of noncommutative gauge theories. 
The only difference is that in the former case it is the transverse space 
to the brane, or rather the compact part of the near horizon geometry
that is being deformed. 
This naturally suggests interpreting the matrix $\mathbf{\Gamma}$ appearing 
in equation (\ref{O33LM}) as some kind of noncommutativity parameter. 
Since noncommutativity in this case is a property of the transverse 
space it manifests itself as a deformation of the matter content of 
the theory.

Before we proceed to the next section where we will further 
clarify this point, we would like to make some final remarks about 
the applicability of the solution generating transformations
illustrated above. Despite the fact that this method has had a rather 
remarkable set of applications so far its utility is unfortunately restricted 
to the following conditions. 
First of all, the directions one wants to introduce fluxes --- or
equivalently noncommutativity --- should be isometry directions realized 
geometrically, meaning as shift symmetries of the metric 
\cite{GiveonRocek}.
In addition, the noncommutativity matrix should have constant entries. 
Expressed in a more precise manner this means that there should exist 
a coordinate system where the noncommutativity is reduced to
a constant along isometry directions of the metric.

As an example of this, let us consider the Melvin Twist gauge theory.
This has been studied in \cite{Alishahihaetal05}\cite{HashimotoThomas04}\cite{HashimotoThomas05}.
The relevant noncommutativity parameter can be written 
in cartesian coordinates as \footnote{Here we consider the case of 
a non compact direction $x_{3}$ in contrast to the most widely
used case.}:

\be\label{Melvincr}
[x_{2},x_{3}]=i b x_{1},\quad [x_{3},x_{1}]=i b x_{2} \quad\text{and}\quad [x_{1},x_{2}]=0
\ee
but in polar coordinates on the $(x_{1},x_{2})$--plane it becomes:

\be
[\rho,\theta ]=0,\quad [\rho,x_{3}]=0,\quad \text{and}\quad[\theta,x_{3}]=i b
\ee
In these coordinates ($\frac{\partial}{\partial\theta},\frac{\partial}{\partial x_{3}}$) are 
indeed Killing vectors of the flat space metric and therefore the solution
generating technique is applicable.

In general it seems reasonable to expect that given a noncommutativity parameter, the following
two conditions should hold for a coordinate system to exist in which
$\nc^{ij}$ is reduced to a constant matrix:


\be\label{noncommutativityconditions}
\left.
\begin{aligned}
\p_{i}\nc^{ij}&=0\\
\nc^{il}\p_{l}\nc^{jk}+\nc^{kl}\p_{l}\nc^{ij}+\nc^{jl}\p_{l}\nc^{ki}&=0 
\end{aligned}
\right\}
\quad \Rightarrow \quad T^{[ijk]}=\p_{l}(\nc^{l[i}\nc^{jk]})=0 
\ee 
Although neither have we been able to find a proof of this
nor have we come accross a proof of it in the literature,
we find that it is natural to think of the second (associativity) 
condition in analogy with the vanishing of the Nijenhus tensor 
condition for an almost complex structure \footnote{It may thus be 
interesting to formulate generalized complex geometry from the point of view of open strings.}. 
We thus understand (\ref{noncommutativityconditions}) as ensuring that one can always find a local
coordinate system in order to put $\nc^{ij}$ in a constant form.
Then, the first condition in (\ref{noncommutativityconditions})
can be read as the possibility of extending the local
coordinates to global ones.\footnote{This is actually not true for the
two--dimensional case, which is particularly simple. For instance,
all noncommutative deformations are also associative ones.} 

We would like to conclude this section by stressing once more
that (\ref{noncommutativityconditions}) cannot be seen as the requirement
for the solution generating transformation to work since
there is no way to make sure that the coordinate transformation
employed to bring $\nc^{ij}$ into a constant form will not 
spoil the shift symmetries present in the metric.
One example of this is the nongeometric background also referred to
as the Q--space in the literature \cite{Sheltonetal}\cite{EllwoodHashimoto}\cite{Loweetal}.
The relevant noncommutativity parameter in this case is:
\be
[x_{1},x_{2}]=i b x_{3},\quad [x_{1},x_{3}]=[x_{2},x_{3}]=0
\ee
While it is obvious from the discussion above that $\nc^{ij}$
can be reduced to a constant, the coordinate transformation that 
makes this possible is \cite{Loweetal}: 
$x_{1}\rightarrow y_{1} y_{3}, x_{2}\rightarrow y_{2},x_{3}\rightarrow y_{3}$
and in these coordinates the metric looks like:
\be
ds^{2}=-dt^2+(y_{1} dy_{3}+y_{3} dy_{1})^2+dy_{2}^2+dy_{3}^2
\ee 
Indeed it has not been possible to embed this noncommutative deformation
of flat space directly into string theory. It nevertheless naturally emerges
when a D3--brane probe is immersed in the background of smeared NS5--branes.


\section{$\beta$--deformations and noncommutativity}\label{BetaNC}

The aim of this section is to establish a precise relation
between transverse space noncommutativity and $\beta$--deformations 
of $\Ncal=4$ SYM. In general the connection between marginal deformations and noncommutativity
is not new. A study of the moduli space clearly points into this direction ---
a thorough analysis can be found in \cite{Berensteinetal00}\cite{Berensteinetal0006}\cite{DoreyHollowood}\cite{Benini}.
The F--term constraints for instance read:
\be\label{Fterms1}
\Phi^{I} \Phi^{J}=q \Phi^{J} \Phi^{I}, \quad \ol{\Phi}^{\bar{I}} \ol{\Phi}^{\bar{J}}=q \ol{\Phi}^{\bar{J}} \ol{\Phi}^{\bar{I}} \quad \text{where $q=e^{2 i\beta}$ and I,J are cyclically ordered.}
\ee 
and $\Phi^{I}$ here indicate the first components of the corresponding superfields. 
These equations are usually understood to represent the space where the D--branes can move. 
For small enough deformations we can interpret the eigenvalues of these matrices as coordinates
parametrizing the transverse space to the worldvolume of the D3--brane.
The eignevalues should however now be thought of as noncommuting numbers 
according to equation (\ref{Fterms1}). If we denote the coordinates 
of the moduli space as $(z^{I},\ol{z}^{\bar{I}})$ with $I,\bar{I}=1,2,3$ we have that:
\be\label{nchol}
z^{I} z^{J}=q z^{J} z^{I}, \quad \ol{z}^{\bar{I}} \ol{z}^{\bar{J}}=q \ol{z}^{\bar{J}} \ol{z}^{\bar{I}} \quad\text{with I,J cyclically ordered}.
\ee
Later on, it will become clear that a noncommutative interpretation is meaningful 
only when $\beta \in \mathbb{R}$. Henceforth we replace $\beta$ with $\gamma$ in 
order to avoid confusion and to be consistent with existing notations in the literature.

As it was mentioned in the previous section we can identify the prescription           
of \cite{LuninMaldacena} with the one used within the context of noncommutative 
gauge theories so long as matrix $\mathbf{\Gamma}$ appearing in equation (\ref{O33LM}) 
is the noncommutativity matrix associated to the deformation of the 
transverse space. Therefore, our main objective here is to construct a noncommutativity 
matrix, or rather a contravariant antisymmetric tensor field $\nc^{IJ}$ to describe the 
deformed space. A natural way to define it is through the commutation relations 
implied by (\ref{nchol}). That is:
\be\label{ncnc}
[z^{I},z^{J}]=i 2 e^{i \gamma} \sin{\gamma} z^{I} z^{J} \quad
[\overline{z}^{I},\overline{z}^{J}]=i 2 e^{i \gamma} \sin{\gamma} \ol{z}^{\bar{I}} \ol{z}^{\bar{J}}
\ee
Clearly such a definition would require a whole different notion of differential geometry since 
the noncommutativity parameter is position dependent and the coordinates themselves are now 
nonocommuting objects. We circumvent this by implementing an alternative procedure. As mentioned in 
the previous section one can replace noncommuting coordinates with commuting ones by  
defining a star product between them. 
In general, constructing an appropriate star product can be an 
equally formidable task as dealing with noncommuting variables.
In this case however a natural proposal was set forth in \cite{LuninMaldacena}.
Specifically, the authors of \cite{LuninMaldacena} suggested:
\be\label{star}
f\ast g=f
e^{i \pi\beta \left(\overleftarrow{Q_{1}}\overrightarrow{Q_{2}}-\overleftarrow{Q_{2}}\overrightarrow{Q_{2}}\right)}
g
\ee
where $f,g$ belong to the set of chiral/antichiral multiplets of the theory 
and $Q_{1,2}$ are the global U(1) charges associated with these fields (see equation (\ref{U1})). 
This proposal was subsequently used \cite{Khoze} in order to rewrite the 
component Lagrangian of the $\beta$--deformed gauge theory as 
the $\Ncal=4$ SYM Lagrangian with the product of matter fields 
now replaced by the above star product. This enabled the author of 
\cite{Khoze} to show that all the amplitudes in the planar limit of 
the deformed theory with $\beta \in \mathbb{R}$ are proportional to
their $\Ncal=4$ counterparts. 
Note that the star here is not explicitly written in terms of derivatives/operators
acting on the fields $(f,g)$. Knowledge of the product in this
form however will be sufficient for the purposes of this letter.

In what follows we will use equation (\ref{star}) in 
order to write down a noncommutativity matrix and compare it 
with (\ref{O33LM}). Then we will discuss ways to derive the
appropriate $\nc^{ij}$ without prior knowledge of the star product.
We therefore define the noncommutativity parameter through the following relations:
\be\label{ncrelations}
\left.
\begin{matrix}
[z^{I},z^{J}]_{\ast}&=\left(z^{I}\ast z^{J}-z^{J}\ast z^{I}\right)=i \nc^{IJ}\\
[\overline{z}^{\bar{I}},\overline{z}^{\bar{J}}]_{\ast}&=\left(\ol{z}^{\bar{I}}\ast \ol{z}^{\bar{J}}-\ol{z}^{\bar{J}}\ast \ol{z}^{\bar{I}}\right)=i \nc^{\bar{I}\bar{J}}\\
[z^{I},\overline{z}^{\bar{J}}]_{\ast}&=\left(z^{I}\ast \ol{z}^{\bar{J}}-\ol{z}^{\bar{J}}\ast z^{I}\right)=i \nc^{I\bar{J}}\\
\end{matrix}
\right\}\quad\quad\Rightarrow\quad\quad
\begin{matrix}
\nc^{IJ}&=2 \sin{\gamma} z^{I}z^{J}\\
\nc^{\bar{I}\bar{J}}&=2 \sin{\gamma} \overline{z}^{\bar{I}}\overline{z}^{\bar{J}}\\
\nc^{I\bar{J}}& =-2 \sin{\gamma}z^{I}\overline{z}^{\bar{J}}\\
\end{matrix}
\ee
with $(I,J)$ cyclically ordered. Setting $a\equiv 2\sin{\gamma}$ and writting this in matrix notation, we obtain:
\be\label{ncbetamatrix} \nc=a \begin{pmatrix} 0&
z_{1}z_{2}&-z_{1}z_{3}&0& -z_{1}\ol{z}_{2}&
z_{1}\ol{z}_{3}\\
-z_{1}z_{2}&0& z_{2}z_{3}&
\ol{z}_{1}z_{2}&0&- z_{2}\ol{z}_{3}\\
 z_{3}z_{1}&- z_{2}z_{3}&0&-
\ol{z}_{1}z_{3}&\ol{z}_{2}z_{3}&0\\
0&-\ol{z}_{1}z_{2}&\ol{z}_{1}z_{3}&0&
\ol{z}_{1}\ol{z}_{2}&-\ol{z}_{1}\ol{z}_{3}\\
z_{1}\ol{z}_{2}&0&-\ol{z}_{2}z_{3}&-
\ol{z}_{1}\ol{z}_{2}&0&\ol{z}_{2}\ol{z}_{3}\\
-\ol{z}_{3}z_{1}&\ol{z}_{3}z_{2}&0& \ol{z}_{1}\ol{z}_{3}&-
\ol{z}_{2}\ol{z}_{3}&0
\end{pmatrix}\ee
Clearly, the result obtained above is not exactly a satisfactory one. 
Despite the fact that we managed to describe the deformation of the transverse 
space in a noncommutative way, the associated noncommutativity matrix $\nc$ is both
position dependent and six dimensional. It does not therefore in any sense 
resemble to matrix $\mathbf{\Gamma}$ of equation (\ref{O33LM}). An additional
interesting but perhaps perplexing feature of $\nc$ is that it is not
a purely holomorphic/antiholomorphic matrix as we might have expected
from the F--term constraints. We will return to this point later in this
section after we outline a more general prescription of identifying
the appropriate $\nc^{ij}$.

Let us however proceed to make a coordinate transformation on (\ref{ncbetamatrix}).
Since $\nc^{IJ}$ thus defined is a contravariant tensor we have no 
trouble doing so. In other words we know that when changing coordinates 
from $\{x^{i}\}$ to $\{x'^{i'}\}$, the noncommutativity parameter transforms 
as:
\be\label{transformationrule}
\nc^{i'j'}=\frac{\p x'^{i'}}{\p x^{i}}\frac{\p x'^{j'}}{\p x^{j}}\nc^{ij}
\ee
Here, we chose to rewrite $\nc^{IJ}$ in spherical coordinates $(r,\alpha,\theta,\phi_{1},\phi_{2},\phi_{3})$ 
defined through: 
\be\begin{split}
z_{1}&=r \cos{\alpha} \mathrm{e}^{i \phi_{1}}, \quad z_{2}=r \sin{\alpha}\sin{\theta} \mathrm{e}^{i \phi_{2}}, \quad
z_{3}=r \sin{\alpha}\cos{\theta} \mathrm{e}^{i \phi_{3}}\\
\ol{z}_{1}&=r \cos{\alpha} \mathrm{e}^{-i \phi_{1}}, \quad \ol{z}_{2}=r \sin{\alpha}\sin{\theta} \mathrm{e}^{-i \phi_{2}}, \quad \ol{z}_{3}=r \sin{\alpha}\cos{\theta} \mathrm{e}^{-i \phi_{3}}
\end{split}
\ee
Note that in these coordinates we should be careful to define if possible the parameter 
$\gamma$ of our matrix so as to have $\nc\in\mathbb{R}$.
Only then can $\nc$ be interpreted as a noncommutativity parameter in the
usual sense.
Applying (\ref{transformationrule}) to (\ref{ncbetamatrix}) we obtain 
in matrix notation:
\be\label{LMThetaPolar} \nc=\begin{pmatrix}
0&0&0&0&0&0\\
0&0&0&0&0&0\\
0&0&0&0&0&0\\
0&0&0&0&-a&a\\
0&0&0&a&0&-a\\
0&0&0&-a&a&0
\end{pmatrix}\ee
and we immediately see that we can indeed think of $\nc$ as a noncommutativity matrix only
when $a\in\mathbb{R}$. More importantly, from equation (\ref{LMThetaPolar}) it is 
clear that we can reduce $\nc$ to the $3\times 3$ matrix denoted as $\mathbf{\Gamma}$ in section \ref{LM}.
\footnote{We can actually reduce $\nc^{ij}$ even further using coordinates: 
$\psi=\frac{1}{3}\sum_{i=1}^3 \phi_{i},\sigma_{1}=\frac{1}{3}(\phi_{2}+\phi_{3}-2 \phi_{1}),\sigma_{2}=\frac{1}{3}(\phi_{1}+\phi_{3}-2 \phi_{2})$.
In this parametrization $\psi$ denotes the U(1) circle associated with the R--symmetry of 
the original background and $\nc^{ij}$ reads:
$\nc=
\left(\begin{smallmatrix}
0&0&0\\
0&0&-a\\
0&a&0\\ 
\end{smallmatrix}\right)$. It is then obvious that the solution generating transformation 
does not act on the $U(1)_{R}$ therefore preserving $\Ncal=1$ supersymmetry.}. 
The only difference is that now the deformation parameter $\gamma$ of the gauge theory is 
replaced by $a=2 \sin\gamma$. Recall however, that the Lunin--Maldacena solution (\ref{LMsolution}) 
has small curvature only when: $\gamma R \ll 1$ and $R\gg 1$. Then $b\simeq 2\gamma$ and the
solutions generated by using either $\Gamma$ or $\nc$ are basically equivalent.
Yet we find it interesting that the periodicity of the parameter $\gamma$ is manifest 
in this description. Nonetheless, note that this is not quite the correct periodicity condition. 
Our result is periodic when $\gamma \rightarrow \gamma+2\pi$ whereas from (\ref{Fterms1}) we expect: 
$\gamma \rightarrow \gamma+\pi$. The reason for this discrepancy lies in equation (\ref{ncrelations}). 
Indeed, the two ways of defining deformed commutators, one in terms of commuting variables multiplied 
with a star product and the other in terms 
of noncommuting ones, are only strictly equivalent when the commutation relations
are c--numbers. 
"Comparing" equations (\ref{ncrelations}) and (\ref{ncnc}) in this case we see that there 
is a a phase difference between the parameters entering the two definitions. 
The absence of this phase in (\ref{ncrelations}) is responsible for the discrepancy 
in periodicity. Nevertheless, the star product gives a more natural way to think of $\nc^{ij}$ 
as a contravariant antisymmetric tensor thus having well defined transformation
properties a change of coordinates. 

Suppose now that no precise definition of a star product between the superfields 
of the theory was known. Would we be able to construct the noncommutativity matrix and 
therefore find the gravity dual of the $\beta$--deformed gauge theory? A glance at the 
superpotential of the theory would naturally lead us to define:
\be
\nc^{IJ}=2 \sin{\gamma} z^{I}z^{J}\quad\text{and}\quad
\nc^{\bar{I}\bar{J}}=2 \sin{\gamma} \overline{z}^{\bar{I}}\overline{z}^{\bar{J}}
\ee
and therefore correctly guess the purely holomorphic and purely
antiholomorphic parts of $\nc^{ij}$. What about the other parts though?
We can actually constrain the form of $\nc^{I\bar{J}}$ by the following
requirements:

\begin{itemize}
\item Definite Reality Properties.

In order to be able to describe the deformation in noncommutative terms
we should define the parameters appearing in $\nc^{ij}$ so as to have
a matrix with real entries after going to real coordinates.
\item Symmetries.

Since we expect the global symmetries of the Langrangian to be preserved
in the strong coupling limit as well, we should ensure that the noncommutativity
matrix respects those symmetries. This is true as long as \cite{Tamassia}:

\be\label{ncsymmetries}
[z^{I},z^{J}]=i \nc^{IJ}(z)\xrightarrow{z\rightarrow z'} 
[z'^{I},z'^{J}]=i \nc^{IJ}(z')
\ee

Note that this is precisely analogous to the condition for a certain symmetry
to be an isometry of the metric. 
Assuming that $\nc^{I\bar{J}}$ is quadratic (\ref{ncsymmetries}) implies that
up to a sign there exist only two possibilities:
$\nc^{I\bar{J}}=0$ or $\nc^{I\bar{J}}=z^{I}\overline{z}^{\bar{J}}$.
\item Marginality condition.                            

According to the usual reasoning of AdS/CFT, marginal defromations should
be described by AdS geometries with different compact pieces. This suggests
that when the noncommutativity parameter is transformed in shperical coordinates,
it should be independent of and have no components along the radial direction of AdS.  
In other words, $\frac{\p\nc^{a_{i}a_{j}}}{\p r}=0$ where $a_{i}$ are angular
variables parametrizing the five sphere and $\nc^{r a_{i}}=0$. This last requirement
completely determines the form of $\nc^{I\bar{J}}$ to be the one appearing
in (\ref{ncrelations}).
\end{itemize} 

We see as remarkable as it may seem that there exists a \emph{unique} 
noncommutativity matrix which respects the above conditions. 
Stated differently, simple gauge theory data and elementary notions 
from the AdS/CFT correspondence, made it possible to fully
determine the form of $\nc^{ij}$.
We thus want to understand this matrix as a way of
encoding the deformation of the transverse space or in other
words, the moduli space of the gauge theory --- at least 
insofar as information relevant to the gauge/gravity duality
in the large N limit is concerned. 
Indeed given the F--term constraints we seem to have extracted 
information coming from the D--terms. We can convince ourselves 
of this with the following observation.  
Recall that the $\beta$--deformation of $\Ncal=4$ SYM
is exactly marginal and that the deformation enters only in the 
superpotential of the theory. This means that we wish not to 
deform the D--terms in the Lagrangian. Note however that we
can write the D--terms of the $\Ncal=4$ theory as:
\be\label{Dterms}
\Tr[\Phi_{I},\tilde{\Phi}^{I}][\Phi_{J},\tilde{\Phi}^{J}]=
\Tr[\Phi_{I},\Phi_{J}][\tilde{\Phi}^{I},\tilde{\Phi}^{J}]+
\Tr[\Phi_{I},\tilde{\Phi}^{J}][\Phi_{J},\tilde{\Phi}^{I}]
\ee
The first term on the right hand side of equation (\ref{Dterms})
is precisely the contribution to the potential coming from
the F--terms. We then deduce that if we wish to retain the 
D--terms unaffected by the deformation of the F--term commutator 
we must induce an appropriate deformation on the commutator between
holomorphic and antiholomorphic fields as well. Surprisingly
enough, the reasoning outlined above seems to have granted us
this exact piece of information.

It is now evident that we can identify the Lunin--Maldacena generating solution 
technique with the method employed in the case of noncommutative gauge theories \cite{Bermanetal}.
The noncommutative data in this context are basically given to us
from the gauge theory Lagrangian. This is quite natural since the deformations we 
are dealing with are exactly marginal. It is worth pointing out here that combined with the knowledge of the gravity dual of the parent $\Ncal=4$ theory, these data made it possible to find the gravity solution dual to the deformed theory.
Unfortunately, this is not as general a statement as it may seem since
the particular method employed was applicable only because there existed a 
coordinate system in which $\nc^{IJ}$ was reduced to a constant and 
along isometry directions of the metric. In a forthcoming letter 
\cite{Kulaxizi0611} we will nevertheless be able to extract some information 
on the gravity duals of the marginally deformed
$\Ncal=4$ theory when the parameter $\rho$ in (\ref{Superpotential}) 
is different than zero.



\section{Applications and New Backgrounds}\label{Extensions}

In the previous section we were able to associate a specific noncommutativity matrix to
the $\beta$--deformed gauge theory. We found that indeed there exists a coordinate
system for which $\nc^{ij}$ is position independent and lies along U(1) isometries of 
the transverse space metric as well as of the $S^{5}$. Identifying the solution
generating transforms of \cite{LuninMaldacena} and \cite{MaldacenaRusso} was then  
a straightforward task. This result naturally opens up two main directions for 
further study --- the first one pertaining to noncommutative gauge theories and 
the second to deformations of $\Ncal=4$ SYM. In what follows we will try to touch upon 
several questions arising in both these cases.

\subsection{Noncommutative gauge theories.}

The most direct application of the ideas discussed so far is to consider the Lunin--Maldacena 
prescription in order to obtain the gravity duals of noncommutative gauge theories with $\beta$--type 
noncommutativity
\footnote{Similar considerations in the context of the Maldacena--Nunez background 
appeared in \cite{GursoyNunez}\cite{Mateosetal}.}.
This simply means that we wish to think of $\nc^{ij}$ or rather $\mathbf{\Gamma}$ of 
(\ref{O33LM}) as 
a noncommutativity matrix along the worldvolume of the D3--brane
\footnote{Obviously the same procedure can be applied to all branes in 
a fashion similar to \cite{Itzhakietal}\cite{Alishahihaetal99}\cite{Alishahihaetal05}.}.
Provided a decoupling limit exists \footnote{One can actually check this by either calculating the
graviton absorption cross--section or the potential that gravitons feel due to the presence of
the D--brane \cite{Alishahihaetal00}.}, 
we can use the solution generating 
technique reviewed in section \ref{LM}, to either deform the p--brane solution itself, 
or the near horizon geometry directly. For reasons of uniformity, we decided to adhere to 
the latter prescription in what follows.
In four dimensional Euclidean space, $\nc^{ij}$ can be written in complex coordinates as: 
\be\label{gammanc}
\begin{split}
[z_{i},z_{j}]=&i b z_{i} z_{j},\quad [\ol{z}_{i},\ol{z}_{j}]=i b \ol{z}_{i} \ol{z}_{j}, \quad
[z_{i},\ol{z}_{j}]=-i b z_{i} \ol{z}_{j}\\ 
&\text{for \quad $i<j$ \quad and \quad i,j=1,2}
\end{split}
\ee
As we already saw in the previous section transforming to polar coordinates yields
a constant noncommutativity parameter along the two--torus:
\be
[\phi_{1},\phi_{2}]=i b,\quad [\rho_{1},\rho_{2}]=[\rho_{i},\phi_{j}]=0 \quad i,j=1,2
\ee
Constructing a matrix out of these relations is a fairly obvious step which leads us to 
matrix $\mathbf{\Gamma}$ appearing in (\ref{O22}). We can therefore directly apply the 
associated T--duality transform (\ref{O22}) on the \AdS geometry. The relevant part of 
the background matrix is: 
\be E=u^2 R^2\begin{pmatrix}
\rho_{1}^2&0\\
0&\rho_{2}^2\\
\end{pmatrix}
\ee
and substituting into (\ref{NSFieldsTransform}) we find:
\be\label{betancdual}
\begin{split}
ds^2_{str}&=ds_{\widetilde{AdS}}^2+ds_{S^{5}}^2,\quad \text{where} \quad 
ds_{\widetilde{AdS}}^2=u^2 R^2(d\rho_{1}^2+d\rho_{2}^2+ G (\rho_{1}^2 d\phi_{1}^2+ \rho_{2}^2 d\phi_{2}^2))\\
B&= \hb R^2 G \rho_{1}^2\rho_{2}^2 u^4 d\phi_{1}\wedge d\phi_{2} , \quad  e^{2\Phi}=G e^{2 \Phi_{0}}\\ 
G&=\frac{1}{1+\hb^2 \rho_{1}^2 \rho_{2}^2 u^4}, \quad \hb =R^2 b\\
F_{3}&=-3(4 \pi\mathrm{N}) b u^3 \rho_{1} \rho_{2} d\rho_{1}\wedge d\rho_{2}\wedge du,\quad
F_{5}=4 \pi \mathrm{N}(\omega_{\widetilde{AdS}}+\omega_{S^{5}})
\end{split}
\ee
with the RR--fields computed using the T--duality rules of 
\cite{Sundell}\cite{Bergshoeffetal}\cite{Cveticetal}\cite{Fukumaetal}\cite{Hassan}.
Note here that the effect of noncommutativity is important for large radial directions
but negligible for small ones. The same behaviour has been observed in the case 
of the Melvin Universe \cite{HashimotoThomas05}\cite{Alishahihaetal05}.
It seems natural therefore to expect that manifestations of this spatial 
nonhomogeneity will be similar to those described in \cite{HashimotoThomas05}.
It would be interesting for this purpose to explore the instanton, monopole and vortex 
solutions of the theory. In the Melvin--twist gauge theory the corresponding analysis 
showed \cite{HashimotoThomas05} that although the length of the magnetic monopole is 
position dependent, its mass agrees with the ordinary SYM monopole solution. 
It is plausible that study of the $\beta$--type noncommutative gauge theory
along these lines will lead to analogous results.
In addition, it is important to investigate the stability properties of the above
solution, since the background in question may generically break supersymmetry 
 (see e.g. \cite{Russo} \cite{Spradlinetal} for a discussion on this point). 
We would like now to proceed and consider the same type of deformation in Lorentzian signature but 
before doing so, let us make a few remarks regarding the action of the gauge theory dual to (\ref{betancdual}).

Clearly, knowledge of an appropriate star product is more often than not, necessary in 
order to specify the action that describes a noncommutative gauge theory. In the case illustrated 
above, $\nc^{ij}$ is position dependent and it is then known that a suitable product is the one 
defined by Kontsevich in \cite{Kontsevich}. 
Naively one would then think that the action of the gauge theory is obtained by simply 
replacing the ordinary product of functions with the star product. 
The latter product is however not compatible with the Leibnitz rule so that 
one should actually employ what is referred to as the "frame formalism" introduced in \cite{BehrSykora}. 
Alternatively, one can take advantage of the fact that $\nc^{ij}$ is constant in polar 
coordinates and specify a Moyal--like product of functions. The precise mapping between 
this product and the one defined by Kontsevich should then be found, which would however 
\emph{not} be the result of a simple change of coordinates. This procedure has been carried out 
explicitly in a number of cases \cite{Cerchiai}\cite{HashimotoThomas05}\cite{Alishahihaetal05} and 
we refer the reader to these papers for details.

Let us now move on to consider the $\beta$--type deformation on a four--dimensional spacetime 
with Lorentzian signature. Performing a wick rotation according to $z\rightarrow i x^{+}, \ol{z}\rightarrow i x^{-}$ 
along with $b\rightarrow i b$ we can write the commutation relations of equation 
(\ref{gammanc}) as:
\be
\begin{split}
[x^{+},z]=&i b x^{+} z,\quad [x^{-},\ol{z}]=i b x^{-} \ol{z}, \quad
[x^{+},\ol{z}]=i b x^{+} \ol{z} \quad [x^{-},z]=i b x^{-} z\quad \text{and}\quad [z,\ol{z}]=[x^{+},x^{-}]=0\\ 
\end{split}
\ee
We therefore see that in this case we have to deal with a 
temporal noncommutativity parameter. In general, field theories 
on spaces with time--like noncommutativity $\nc^{0i}\neq 0$ are acausal 
\cite{Seibergetal}\cite{Gopakumaretal}
whereas their quantum counterparts are not unitary. A decoupled \emph{field theory}
limit for D--branes in this case does not exist. It was however found
in \cite{Seibergetal}\cite{Gopakumaretal} that a scaling limit where the closed string sector 
can be separated from the open string one is indeed possible. Massive open strings 
do not decouple in this limit which thus defines a noncommutative open
string theory (NCOS) rather than a field theory. Several aspects of these
NCOS theories are explored in \cite{Gubseretal0009}\cite{BarbonRabinovici}
\cite{RussoSheikhJabbari}\cite{Niarchos00}\cite{Sahakian01}\cite{FriessGubser}.

The precise analysis of which types of noncommutativity lead to unitary theories 
and which not, was carried out in \cite{Aharonyetal00} along the lines of \cite{GomisMehen}. 
There it was shown that a necessary condition for unitarity is that the following 
inner product between external momenta is positive definite:
\be\label{unitarity}
p\diamond p\equiv -p_{\mu}\nc^{\mu\sigma}\Gcal_{\sigma\tau}\nc^{\tau\nu}p_{\nu} >0
\ee
where $\Gcal$ is the background metric for the open strings
and the corresponding field theory.
Let us therefore evaluate this quantity for the $\beta$--like noncommutativity
under consideration here. It is easier if we first perform a coordinate
transformation to go from coordinates $(t, x_{1},x_{2},x_{3})$ 
to $(\tau,\theta,r,\phi)$ defined as: $t=\tau \cosh{\theta}$, $x_{1}=\tau \sinh{\theta}$,
$x_{2}=r\cos{\phi}$ and $x_{3}=r \sin{\phi}$. Here $\tau \epsilon (-\infty,\infty)$,$r \epsilon [0,\infty)$ whereas $\theta$ can be chosen compact or non compact. 
This transformation will bring the commutation relations to the form
\footnote{These coordinates cover half of $\mathbb{R}^{1,3}$\cite{Spradlinetal}.}:
\be\label{gammancL}
[\theta,\phi]=i b \quad\text{and}\quad [\tau,r]=[r,\phi]=[\tau,\phi]=[\tau,\theta]=[r,\theta]=0
\ee
and substituting into (\ref{unitarity}) we obtain: $p\diamond p=b^2 (p_{\theta}^2 r^2+p_{\phi}^2 \tau^2)$
which is clearly positive definite. Can we therefore deduce that the $\beta$--type noncommutative
deformation describes a unitary field theory? To be precise, the unitarity requirement of (\ref{unitarity}) 
is proven for a position independent noncommutativity parameter turned on in flat space. 
In our case, as soon as we go to a reference frame where $\nc$ is constant, the corresponding
spacetime exhibits a time--dependent behaviour. It is therefore ambiguous what the meaning of
unitarity is in this context. 

It may be interesting however to address these issues through the dual gravity description of this theory. 
Let us therefore apply the T--duality transformation rules in order to construct this background. 
Alternatively, we can wick rotate the Euclidean solution of equation (\ref{betancdual}) according to
$\rho_{1}\rightarrow i \tau,\phi_{1} \rightarrow i \theta, b\rightarrow i b$. Either way we obtain 
\footnote{The resulting background appears to be well defined due to the particular nature
of the wick rotation employed. This fact does not seem to indicate the need for another kind of scaling 
limit as usual in the dual description of NCOS theories. We would therefore naively expect that indeed this 
supergravity solution is dual to a field theory.}:
\be\label{betancdualtimedependent}
\begin{split}
ds^2_{str}&=ds_{\widetilde{AdS}}^2+ds_{S^{5}}^2,\quad \text{where} \quad 
ds_{\widetilde{AdS}}^2=u^2 R^2(-d\tau^2+ dr^2+ G (\tau^2 d\theta^2+ r^2 d\phi^2))\\
B&= \hb R^2 G \tau^2 r^2 u^4 d\theta\wedge d\phi , \quad  e^{2\Phi}=G e^{2 \Phi_{0}}\\ 
G&=\frac{1}{1+\hb^2 \tau^2 r^2 u^4}, \quad \hb =R^2 b\\
F_{3}&=-3 (4 \pi\mathrm{N}) b u^3 \tau r d\tau\wedge dr\wedge du,\quad
F_{5}=4 \pi \mathrm{N}(\omega_{\widetilde{AdS}}+\omega_{S^{5}})
\end{split}
\ee
Note again that equation (\ref{betancdualtimedependent}) defines a time dependent background
dual to a noncommutative theory which can be thought of as living either in flat space with 
temporal time--dependent noncommutativity parameter or in a time--dependent background which 
is noncommutative only along some of the spatial directions. 
Similar time--dependent configurations were explored in 
\cite{HashimotoSethi}\cite{AlishahihaParvizi}\cite{Liuetal0204}\cite{Liuetal0206}. 
For the case of compact $\theta$ with $\theta \sim\theta+2\pi$ and rational parameter $\beta$,
the gravity solution (\ref{betancdualtimedependent}) corresponds to the near horizon geometry 
of a D3--brane immersed in a time--dependent background that admits an orbifold description
\cite{Spradlinetal}\cite{Tai}\cite{GregoryHarvey}\cite{Headrick}. The latter deformation of flat space 
can be recovered from flat space with the same technique \cite{Spradlinetal}: 
\be\label{flatbetatimebg}
\begin{split}
ds^2&=-d\tau^2+ dr^2+ \frac{\tau^2}{1+b^2 \tau^2 r^2}d\theta^2 +\frac{r^2}{1+b^2\tau^2 r^2}d\phi^2\\
e^{2\Phi}&=\frac{1}{1+b^2 \tau^2 r^2}\\
B&=-\frac{b\tau^2 r^2}{1+b^2 \tau^2 r^2} d\theta\wedge d\phi
\end{split}
\ee
The background indicated above presents an interesting time evolution noted in \cite{Spradlinetal}.
In particular, it appears to be periodically changing for the designated choices 
of $\theta$ and $\beta$.  At $\tau=-\infty$ it is described via the orbifold
 $[\mathbb{R}^{1,1}/\mathbb{Z}]_{\Delta=2\pi}\times[\mathbb{C}/\mathbb{Z}_{N}]$ 
(i.e. orbifold by the boost $\Delta=2\pi$) which gradually evolves
to $[\mathbb{R}^{1,1}/\mathbb{Z}]_{\Delta=2\pi}\times \mathbb{C}$ at time $\tau=0$. Then the reverse 
process begins until it reaches the original orbifold description at $\tau=\infty$.
In complete analogy, the spacetime of equation (\ref{betancdualtimedependent})
shows a periodic evolution with the effects of noncommutativity becoming
most important at $\tau=\pm \infty$ but negligible at $\tau=0$ where the geometry
tends to \AdS. 

This completes our discussion of noncommutative gauge theories. 
We have clearly here only alluded to a number of issues regarding these theories
and noncommutative spacetimes in general. It would certainly be of interest 
to explore these issues further in the future.

\subsection{Matter--content deformations of $\Ncal=4$ SYM}

A natural question to ask in this context is whether we can now borrow results pertaining 
to noncommutative gauge theories in order to explore different kinds of (super)potential 
deformations of $\Ncal=4$ SYM. A few cases where the solution generating technique was 
applicable were already mentioned in section \ref{NC}.
Consider for instance the original situation where a constant noncommutativity parameter is turned on. 
Here, we would like to translate this deformation to some kind of transverse space noncommutativity.
If we parametrize our six dimensional space with complex coordinates $(z^{I},\ol{z}^{\bar{I}})$,
we can write:
\be\label{cr}
[z^{I},z^{J}]=i b,\quad [\ol{z}^{\bar{I}},\ol{z}^{\bar{J}}]=i b, 
\quad [z^{I},\ol{z}^{\bar{J}}]=-i b \quad \text{with I,J cyclically ordered}
\ee
We may then associate these commutation relations to a deformation of the gauge theory potential $\Vcal$. 
Since the type of deformations considered in this section may generically 
break supersymmetry we prefer to state the deformation in terms of the potential which of course
may when appropriate be promoted to the superpotential.
Identifying the coordinates $(z^{I},\ol{z}^{\bar{I}})$ with the 
scalar fields of the theory would naturally lead to\footnote{This deformation is only meaningful for gauge 
groups other than SU(N).}:
\be\label{constanttransverse}
\Vcal_{\Ncal=4}=\Tr[\Phi^{I},\Phi^{J}][\Phi^{\bar{I}},\Phi^{\bar{J}}]\rightarrow
\Tr[\Phi^{I},\Phi^{J}][\Phi^{\bar{I}},\Phi^{\bar{J}}]_{\ast}.
\ee 
Here the star product is defined according to (\ref{cr}) as: $\Phi^{I}\ast\Phi^{J}=\Phi^{I}\Phi^{J}+i b$.
In a similar fashion we could relate the noncommutative deformation of the Melvin Universe
which in complex coordinates looks like
\footnote{Here we defined $z_{1}\equiv x_{1}+i x_{2}$ and $z_{2}\equiv x_{3}+i x_{4}$ with $x_{i}$ 
as appearing in equation (\ref{Melvincr}).}:
\be\label{crM}
[z_{1},z_{2}]=-b z_{1},\quad[\ol{z}_{1},\ol{z}_{2}]=b \ol{z}_{1},\quad [z_{1},\ol{z}_{2}]=-b z_{1}\quad[\ol{z}_{1},z_{2}]=b \ol{z}_{1}\quad\text{with all other commutators vanishing}
\ee
to a potential deformation of the same form as in (\ref{constanttransverse}) but with
a different star product as indicated from (\ref{crM}). 

Yet the true story is not as simple as this. These deformations are
not marginal and the theory will generically flow from some UV point to an IR one.
This means that we cannot solely rely on the data given to us from the Lagrangian
of the theory which we can only take to be a valid description near the UV (small b).
Moreover, the precise arguments that helped us construct the noncommutativity matrix
encoding the moduli space in the $\beta$--deformed case are not applicable anymore.
We do not therefore have a means of understanding the commutation relations
between holomorphic and antiholomorphic fields/coordinates despite the fact that
we believe such a cosntruction may be possible in the future. In addition
we do not even know whether a noncommutative description of the 
transverse space will be valid throughout the flow
\footnote{Note however that it is possible to further examine this in certain cases, 
especially when some of the fields can integrate out by considering the theory at 
appropriate energy scales.}.

Nevertheless, we could still expect to find the relevant supergravity solutions
and use that as a means of understanding the precise gauge theory duals.
Unfortunately this is again a difficult task to pursue because the
solution generating technique discussed in this paper in not applicable anymore.
The reason for this lies in the fact that the directions where the noncommutativity
parameter is constant are not isometry directions of the transverse part of 
the D3--brane geometry. One could of course apply the T--duality transform
on flat space. This would give rise to a deformed flat space geometry 
with non--trivial B--field and dilaton turned on, in which once D3--branes are 
immersed and the near horizon limit is taken, would result in the appropriate gravity dual. 
We think that it will be very interesting to explore this point further as well as to study the corresponding gauge theories which we schemmatically described above.

\section{Conclusion}\label{Conclusion}

In this article we established a precise relation between noncommutativity and $\beta$--deformations
of $\Ncal=4$ SYM theory. We first identified a specific matrix within the solution generating 
transform of \cite{LuninMaldacena}\cite{Aybike} which plays the role of noncommutativity parameter 
$\nc^{ij}$ and then showed how it arises from the gauge theory point of view. Moreover, we 
explained that it is possible to fully specify $\nc^{ij}$ by imposing requirements on 
its particular form naturally deduced from the gauge theory and AdS/CFT. We further argued that $\nc^{ij}$
thus constructed encompasses all the relavant information on the moduli space of the gauge theory. 

This hints at an alternative path in exploring deformations of the original AdS/CFT proposal \cite{Maldacena} which
consists in first specifying the associated \emph{open} string parameters and then mapping them to 
the \emph{closed} string ones. Here we investigated the former issue for the particular case of a 
Leigh--Strassler marginal deformation of the $\Ncal=4$ SYM theory. The mapping to the closed degrees 
of freedom in this case was granted to us in the form of T--duality transformation rules.
In a forthcoming publication \cite{Kulaxizi0611} we will combine the basic reasoning set forth in this note 
with an attempt to address the latter issue in a situation where U(1) symmetries are absent
and the T--duality prescription is not applicable. Such is the case for the superpotential deformation of equation 
(\ref{Superpotential}) with $\rho\neq 0$.

There are many possibilities for future work.
A natural and possibly straightforward generalization would be to consider marginal 
deformations of theories with matter fields in the bifundamental, or in other words 
situations with the D--branes sitting at the orbifold fixed point.  
It would furthermore be of interest to extend this formulation if possible to deformations
which are not marginal, mass deformations for instance. Since both cases have been studied in 
alternative ways \cite{PilchWarner}\cite{Halmagyietal}\cite{KlebanovWitten} 
they appear to provide a sound testing ground for the ideas proposed herein.

A curious feature of this approach is that supersymmetry does not play any central role in it.
Indeed the whole discussion so far has solely relied on the commutation relations between the scalar 
fields of the theory. When however supersymmetry is preserved scalars are accompanied by their 
fermionic superpartners and it is obvious that similar (anti)commutation relations will
be obeyed by the fermions alone as well as between the scalars and the fermions of the theory. 
It seems plausible to us that information pertaining to these (anti)commutation relations is hidden 
in the RR sector of the theory \cite{deBoeretal}\cite{OoguriVafa0302}\cite{BerkovitsSeiberg} it would 
therefore be of great importance to study it in a similar fashion. 

As a natural application of the connection between $\beta$--deformations and noncommutativity
in this article we also constructed the gravity duals of noncommutative gauge theories 
with $\beta$--type noncommutativity. As mentioned in the previous section, the corresponding 
backgrounds may generically be unstable. Moreover, in the particular case of Lorentzian signature the 
gravity solution presents an interesting time evolution. Precisely due to these features, a lot of 
interesting questions arise which are only touched upon in this note and certainly deserve deeper study.

In summary, we have presented a concrete realization of noncommutativity in the context of marginal 
deformations of $\Ncal=4$ SYM. We believe this opens up another window into understanding
the AdS/CFT correspondence which we hope to further explore in the future.


\vskip 15mm
\subsubsection*{Acknowledgments}

I am particularly grateful to K. Zoubos for numerous discussions and a careful 
reading of the manuscript. I would also like to thank A. Murugan, A. Parnachev,
 L. Rastelli and R. Ricci for helpful conversations and comments.


\bibliography{paper1refs}

\providecommand{\href}[2]{#2}\begingroup\raggedright\begin{thebibliography}{10}

\bibitem{Maldacena}
J.~Maldacena, {\it The large {N} limit of {S}uperconformal theories and
  supergravity},  {\em Adv. Theor. Math. Phys.} {\bf 2} (1998) 231,
  [\href{http://xxx.lanl.gov/abs/hep-th/9711200}{{\tt hep-th/9711200}}].

\bibitem{Gubseretal98}
S.~S. Gubser, I.~R. Klebanov, and A.~M. Polyakov, {\it Gauge {T}heory
  {C}orrelators from {N}on--{C}ritical {S}tring {T}heory},  {\em Phys. Lett. B}
  {\bf 428} (1998) 105--114,
  [\href{http://xxx.lanl.gov/abs/hep-th/9802109}{{\tt hep-th/9802109}}].

\bibitem{Witten9802}
E.~Witten, {\it Anti {D}e {S}itter {S}pace {A}nd {H}olography},  {\em Adv.
  Theor. Math. Phys.} {\bf 2} (1998) 253--291,
  [\href{http://xxx.lanl.gov/abs/hep-th/9802150}{{\tt hep-th/9802150}}].

\bibitem{LuninMaldacena}
O.~Lunin and J.~Maldacena, {\it Deforming field theories with
  {U}(1)$\times${U}(1) global symmetry and their gravity duals},  {\em JHEP}
  {\bf 0505} (2005) 033, [\href{http://xxx.lanl.gov/abs/hep-th/0502086}{{\tt
  hep-th/0502086}}].

\bibitem{LeighStrassler95}
R.~G. Leigh and M.~J. Strassler, {\it Exactly marginal operators and duality in
  four dimensional {$\Ncal=1$} supersymmetric gauge theory},  {\em Nucl. Phys.
  B} {\bf 447} (1995) 95, [\href{http://xxx.lanl.gov/abs/hep-th/9503121}{{\tt
  hep-th/9503121}}].

\bibitem{Frolovetal0503}
S.~A. Frolov, R.~Roiban, and A.~Tseytlin, {\it Gauge--string duality for
  superconformal deformations of ${\Ncal}=4$ {S}uper {Y}ang--{M}ills theory},
  {\em JHEP} {\bf 0507} (2005) 045,
  [\href{http://xxx.lanl.gov/abs/hep-th/0503192}{{\tt hep-th/0503192}}].

\bibitem{Frolovetal0507}
S.~A. Frolov, R.~Roiban, and A.~Tseytlin, {\it Gauge--string duality for
  (non)supersymmetric deformations of ${\Ncal}=4$ {S}uper {Y}ang--{M}ills
  theory},  {\em Nucl.Phys. B} {\bf 731} (2005) 1--44,
  [\href{http://xxx.lanl.gov/abs/hep-th/0507021}{{\tt hep-th/0507021}}].

\bibitem{ChenKumar}
H.-Y. Chen and S.~P. Kumar, {\it Precision {T}est of {A}d{S}/{CFT} in
  {L}unin--{M}aldacena {B}ackground},  {\em JHEP} {\bf 0603} (2006) 051,
  [\href{http://xxx.lanl.gov/abs/hep-th/0511164}{{\tt hep-th/0511164}}].

\bibitem{Durnfordetal}
C.~Durnford, G.~Georgiou, and V.Khoze, {\it Instanton test of
  non--supersymmetric deformations of the ${A}d{S}_{5}\times {S}^{5}$},  {\em
  JHEP} {\bf 0609} (2006) 005,
  [\href{http://xxx.lanl.gov/abs/hep-th/0606111}{{\tt hep-th/0606111}}].

\bibitem{GeorgiouKhoze}
G.~Georgiou and V.~Khoze, {\it Instanton {C}alculations in the beta--deformed
  {A}d{S}/{CFT} {C}orrespondence},  {\em JHEP} {\bf 0604} (2006) 049,
  [\href{http://xxx.lanl.gov/abs/hep-th/0602141}{{\tt hep-th/0602141}}].

\bibitem{Hernandezetal}
R.~Hernandez, K.~Sfetsos, and D.~Zoakos, {\it Gravity duals for the {C}oulomb
  branch of marginally deformed ${\Ncal}=4$ {Y}ang--{M}ills},  {\em JHEP} {\bf
  0603} (2006) 069, [\href{http://xxx.lanl.gov/abs/hep-th/0510132}{{\tt
  hep-th/0510132}}].

\bibitem{BerensteinCherkis04}
D.~Berenstein and S.~A. Cherkis, {\it Deformations of {$N=4$} {SYM} and
  integrable spin chain model},  {\em Nucl. Phys. B} {\bf 702} (2004) 49--85,
  [\href{http://xxx.lanl.gov/abs/hep-th/0405215}{{\tt hep-th/0405215}}].

\bibitem{BeisertRoiban}
N.~Beisert and R.~Roiban, {\it Beauty and the {T}wist: The {B}ethe {A}nsatz for
  {T}wisted ${\Ncal}=4$ {SYM}},  {\em JHEP} {\bf 0508} (2005) 039,
  [\href{http://xxx.lanl.gov/abs/hep-th/0505187}{{\tt hep-th/0505187}}].

\bibitem{Frolov}
S.~Frolov, {\it Lax {P}air for {S}trings in {L}unin--{M}aldacena {B}ackground},
   {\em JHEP} {\bf 0505} (2005) 069,
  [\href{http://xxx.lanl.gov/abs/hep-th/0503201}{{\tt hep-th/0503201}}].

\bibitem{Khoze}
V.~V. Khoze, {\it Amplitudes in the beta--deformed {C}onformal
  {Y}ang--{M}ills},  {\em JHEP} {\bf 0602} (2006) 040,
  [\href{http://xxx.lanl.gov/abs/hep-th/0512194}{{\tt hep-th/0512194}}].

\bibitem{FreedmanGursoy}
D.~Z. Freedman and U.~Gursoy, {\it Comments on the $\beta$--deformed
  ${\Ncal}=4$ {SYM} {T}heory},  {\em JHEP} {\bf 0511} (2005) 042,
  [\href{http://xxx.lanl.gov/abs/hep-th/0506128}{{\tt hep-th/0506128}}].

\bibitem{KuzenkoTseytlin}
S.~M. Kuzenko and A.~Tseytlin, {\it Effective action of $\beta$--deformed
  ${\Ncal}=4$ {SYM} theory and {A}d{S}/{CFT}},  {\em Phys. Rev. D} {\bf 72}
  (2005) 075005, [\href{http://xxx.lanl.gov/abs/hep-th/0508098}{{\tt
  hep-th/0508098}}].

\bibitem{Rossietal05}
G.~C. Rossi, E.~Sokatchev, and Y.~S. Stanev, {\it New results in the deformed
  ${\Ncal}=4$ {SYM} theory},
  \href{http://xxx.lanl.gov/abs/hep-th/0606284}{{\tt hep-th/0606284}}.

\bibitem{Elmettietal}
F.~Elmetti, A.~Mauri, S.~Penati, A.~Santambrogio, and D.~Zanon, {\it Conformal
  invariance of the planar beta--deformed ${\Ncal}=4$ {SYM} theory requires
  beta real},  \href{http://xxx.lanl.gov/abs/hep-th/0606125}{{\tt
  hep-th/0606125}}.

\bibitem{Rossietal06}
G.~C. Rossi, E.~Sokatchev, and Y.~S. Stanev, {\it On the all--order
  perturbative finiteness of the deformed ${\Ncal}=4$ theory},
  \href{http://xxx.lanl.gov/abs/hep-th/0606284}{{\tt hep-th/0606284}}.

\bibitem{Maurietal}
A.~Mauri, S.~Penati, M.~Pirrone, A.~Santambrogio, and D.~Zanon, {\it On the
  perturbative chiral ring for marginally deformed ${\Ncal}=4$ {SYM} theories},
   {\em JHEP} {\bf 0608} (2006) 072,
  [\href{http://xxx.lanl.gov/abs/hep-th/0605145}{{\tt hep-th/0605145}}].

\bibitem{GursoyNunez}
U.~Gursoy and C.~Nunez, {\it Dipole deformations of ${\Ncal}=1$ {SYM} and
  {S}upergravity backgrounds with {U}(1)$\times${U}(1) global symmetry},  {\em
  Nucl. Phys. B} {\bf 725} (2005) 45--92,
  [\href{http://xxx.lanl.gov/abs/hep-th/0505100}{{\tt hep-th/0505100}}].

\bibitem{AhnVazquezPoritz05}
C.~Ahn and J.~F. Vazquez-Poritz, {\it Marginal {D}eformations with ${U}(1)^{3}$
  {G}lobal {S}ymmetry},  {\em JHEP} {\bf 0507} (2005) 032,
  [\href{http://xxx.lanl.gov/abs/hep-th/0505168}{{\tt hep-th/0505168}}].

\bibitem{AhnVazquezPoritz06}
C.~Ahn and J.~F. Vazquez-Poritz, {\it From {M}arginal {D}eformations to
  {C}onfinement},  {\em JHEP} {\bf 0606} (2006) 061,
  [\href{http://xxx.lanl.gov/abs/hep-th/0603142}{{\tt hep-th/0603142}}].

\bibitem{Rashkovetal}
R.~C. Rashkov, K.~S. Viswanathan, and Y.~Yang, {\it Generalizations of
  {L}unin--{M}aldacena transformation on the ${A}d{S}_{5}\times {S}^{5}$
  background},  {\em Phys. Rev. D} {\bf 72} (2005) 106008,
  [\href{http://xxx.lanl.gov/abs/hep-th/0509058}{{\tt hep-th/0509058}}].

\bibitem{Aybike}
A.~Catal-Ozer, {\it Lunin--{M}aldacena {D}eformations with {T}hree parameters},
   {\em JHEP} {\bf 0602} (2006) 026,
  [\href{http://xxx.lanl.gov/abs/hep-th/0512290}{{\tt hep-th/0512290}}].

\bibitem{Kulaxizi0611}
M.~Kulaxizi, ``Marginal {D}eformations and {O}pen vs. {C}losed {S}tring
  {P}arameters.'' To appear.

\bibitem{GiveonPorratiRabinovici}
A.~Giveon, M.~Porrati, and E.~Rabinovici, {\it Traget {S}pace {D}uality in
  {S}tring {T}heory},  {\em Phys. Rept.} {\bf 244} (1994) 77--202,
  [\href{http://xxx.lanl.gov/abs/hep-th/9401139}{{\tt hep-th/9401139}}].

\bibitem{Sundell}
P.~Sundell, {\it Spin(p+1,p+1) {C}ovariant {D}p--brane {B}ound {S}tates},  {\em
  Int. J. Mod. Phys. A} {\bf 16} (2001) 3025--3040,
  [\href{http://xxx.lanl.gov/abs/hep-th/0011283}{{\tt hep-th/0011283}}].

\bibitem{Bergshoeffetal}
E.~Bergshoeff, C.~M. Hull, and T.~Ortin, {\it Duality in the {T}ype--{II}
  {S}uperstring {E}ffective {A}ction},  {\em Nul. Phys. B} {\bf 451} (1995)
  547--578, [\href{http://xxx.lanl.gov/abs/hep-th/9504081}{{\tt
  hep-th/9504081}}].

\bibitem{Cveticetal}
M.~Cvetic, H.~Lu, C.~N. Pope, and K.~S. Stelle, {\it T--duality in the
  {G}reen--{S}chwarz {F}ormalism and the {M}assless/{M}assive {IIA} {M}ap},
  {\em Nucl. Phys. B} {\bf 573} (2000) 149--176,
  [\href{http://xxx.lanl.gov/abs/hep-th/9907202}{{\tt hep-th/9907202}}].

\bibitem{Fukumaetal}
M.~Fukuma, T.~Oota, and H.~Tanaka, {\it Comments on {T}--dualities of
  {R}amond--{R}amond {P}otentials},  {\em Prog. Theor. Phys.} {\bf 103} (2000)
  425--446, [\href{http://xxx.lanl.gov/abs/hep-th/9907132}{{\tt
  hep-th/9907132}}].

\bibitem{Hassan}
S.~F. Hassan, {\it T--{D}uality, {S}pace--time {S}pinors and {R--R} {F}ields in
  {C}urved {B}ackgrounds},  {\em Nucl. Phys. B} {\bf 568} (2000) 145--161,
  [\href{http://xxx.lanl.gov/abs/hep-th/9907152}{{\tt hep-th/9907152}}].

\bibitem{KlebanovWitten}
I.~Klebanov and E.~Witten, {\it Superconformal {F}ield {T}heory on
  {T}hreebranes at a {C}alabi--{Y}au {S}ingularity},  {\em Nucl. Phys.B} {\bf
  536} (1998) 199--218, [\href{http://xxx.lanl.gov/abs/hep-th/9807080}{{\tt
  hep-th/9807080}}].

\bibitem{DouglasNekrasov}
M.~R. Douglas and N.~N. Nekrasov, {\it Noncommutative {F}ield {T}heory},  {\em
  Rev Mod. Phys.} {\bf 73} (2001) 977--1029,
  [\href{http://xxx.lanl.gov/abs/hep-th/0106048}{{\tt hep-th/0106048}}].

\bibitem{MaldacenaRusso}
J.~Maldacena and J.~Russo, {\it Large {N} limit of {N}on--commutative gauge
  theories},  {\em JHEP} {\bf 9909} (1999) 025,
  [\href{http://xxx.lanl.gov/abs/hep-th/9908134}{{\tt hep-th/9908134}}].

\bibitem{HashimotoItzhaki}
A.~Hashimoto and N.~Itzhaki, {\it Non--commutative {Y}ang--{M}ills and the
  {A}d{S}/{CFT} {C}orrespondence},  {\em Phys. Lett. B} {\bf 465} (1999)
  142--147, [\href{http://xxx.lanl.gov/abs/hep-th/9907166}{{\tt
  hep-th/9907166}}].

\bibitem{Bermanetal}
D.~S. Berman, V.~L. Campos, M.~Cederwall, U.~Gran, H.~Larsson, M.~Nielsen,
  B.~E.~W. Nilsson, and P.~Sundell, {\it Holographic {N}oncommutativity},  {\em
  JHEP} {\bf 0105} (2001) 002,
  [\href{http://xxx.lanl.gov/abs/hep-th/0011282}{{\tt hep-th/0011282}}].

\bibitem{Itzhakietal}
N.~Itzhaki, J.~Maldacena, J.~Sonnenschein, and S.~Yankielowicz, {\it
  {S}upergravity and the {L}arge {N} limit of {T}heories with {S}ixteen
  {S}upercharges},  {\em Phys. Rev. D} {\bf 58} (1998) 046004,
  [\href{http://xxx.lanl.gov/abs/hep-th/9802042}{{\tt hep-th/9802042}}].

\bibitem{Alishahihaetal99}
M.~ALishahiha, Y.~Oz, and M.~Sheikh-Jabbari, {\it Supergravity and {L}arge {N}
  {N}oncommutative {F}ield {T}heories},  {\em JHEP} {\bf 9911} (1999) 007,
  [\href{http://xxx.lanl.gov/abs/hep-th/9909215}{{\tt hep-th/9909215}}].

\bibitem{Alishahihaetal05}
M.~Alishahiha, B.~Safarzadeh, and H.~Yavartanoo, {\it On {S}upergravity
  {S}olutions of {B}ranes in {M}elvin {U}niverses},  {\em JHEP} {\bf 0601}
  (2006) 153, [\href{http://xxx.lanl.gov/abs/hep-th/0512036}{{\tt
  hep-th/0512036}}].

\bibitem{SeibergWitten9908}
N.~Seiberg and E.~Witten, {\it String theory and noncommutative geometry},
  {\em JHEP} {\bf 9909} (1999) 032,
  [\href{http://xxx.lanl.gov/abs/hep-th/9908142}{{\tt hep-th/9908142}}].

\bibitem{GiveonRocek}
A.~Giveon and M.~Rocek, {\it Generalized {D}uality in {C}urved
  {S}tring--{B}ackgrounds},  {\em Nucl. Phys. B} {\bf 380} (1992) 128--146,
  [\href{http://xxx.lanl.gov/abs/hep-th/9112070}{{\tt hep-th/9112070}}].

\bibitem{HashimotoThomas04}
A.~Hashimoto and K.~Thomas, {\it {D}ualities, {T}wists, and {G}auge {T}heories
  with {N}on--constant {N}on--{C}ommutativity},  {\em JHEP} {\bf 0501} (2005)
  033, [\href{http://xxx.lanl.gov/abs/hep-th/0410123}{{\tt hep-th/0410123}}].

\bibitem{HashimotoThomas05}
A.~Hashimoto and K.~Thomas, {\it Non--commutative gauge theory on {D}--branes
  in {M}elvin {U}niverses},  {\em JHEP} {\bf 0601} (2006) 083,
  [\href{http://xxx.lanl.gov/abs/hep-th/0511197}{{\tt hep-th/0511197}}].

\bibitem{Sheltonetal}
J.~Shelton, W.~Taylor, and B.~Wecht, {\it Nongeometric {F}lux
  {C}ompactifications},  {\em JHEP} {\bf 0510} (2005) 085,
  [\href{http://xxx.lanl.gov/abs/hep-th/0508133}{{\tt hep-th/0508133}}].

\bibitem{EllwoodHashimoto}
I.~Ellwood and A.~Hashimoto, {\it Effective descriptions of branes on
  non--geometric tori},  \href{http://xxx.lanl.gov/abs/hep-th/0607135}{{\tt
  hep-th/0607135}}.

\bibitem{Loweetal}
D.~Lowe, H.~Nastase, and S.~Ramgoolam, {\it Massive {IIA} string theory and
  matrix theory compactification},  {\em Nucl. Phys. B} {\bf 667} (2003)
  55--89, [\href{http://xxx.lanl.gov/abs/hep-th/0303173}{{\tt
  hep-th/0303173}}].

\bibitem{Berensteinetal00}
D.~Berenstein, V.~Jejjala, and R.~G. Leigh, {\it Marginal and relevant
  deformations of {$N=4$} field theories and non--commutative moduli spaces of
  vacua},  {\em Nucl. Phys. B} {\bf 589} (2000) 196,
  [\href{http://xxx.lanl.gov/abs/hep-th/0005087}{{\tt hep-th/0005087}}].

\bibitem{Berensteinetal0006}
D.~Berenstein, V.~Jejjala, and R.~Leigh, {\it Non--{C}ommutative {M}oduli
  {S}paces, {D}ielectric {T}ori and {T}--duality},  {\em Phys. Lett. B} {\bf
  493} (2000) 162--168, [\href{http://xxx.lanl.gov/abs/hep-th/0006168}{{\tt
  hep-th/0006168}}].

\bibitem{DoreyHollowood}
N.~Dorey and T.~J. Hollowood, {\it On the {C}oulomb {B}ranch of a {M}arginal
  {D}eformation of ${\Ncal}=4$ {SUSY} {Y}ang--{M}ills},  {\em JHEP} {\bf 0506}
  (2005) 036, [\href{http://xxx.lanl.gov/abs/hep-th/0411163}{{\tt
  hep-th/0411163}}].

\bibitem{Benini}
F.~Benini, {\it The {C}oulomb branch of the {L}eigh--{S}trassler deformation
  and matrix models},  {\em Phys. Lett. B} {\bf 493} (2000) 162--168,
  [\href{http://xxx.lanl.gov/abs/hep-th/0411057}{{\tt hep-th/0411057}}].

\bibitem{Tamassia}
L.~Tamassia, {\it Noncommutative {S}upersymmetric/{I}ntegrable {M}odels and
  {S}tring {T}heory},  {\em Scientifica Acta Quaderni del Dottorato} {\bf XX}
  (2005) [\href{http://xxx.lanl.gov/abs/hep-th/0506064}{{\tt hep-th/0506064}}].
  Ph.D. thesis.

\bibitem{Mateosetal}
T.~Mateos, J.~M. Pons, and P.~Talavera, {\it Supergravity {D}ual of
  {N}oncommutative ${\Ncal}=1$ {SYM}},  {\em Nucl. Phys. B} {\bf 651} (2003)
  291--312, [\href{http://xxx.lanl.gov/abs/hep-th/0209150}{{\tt
  hep-th/0209150}}].

\bibitem{Alishahihaetal00}
M.~Alishahiha, H.~Ita, and Y.~Oz, {\it Graviton {S}cattering on {D}6 {B}ranes
  with {B} {F}ields},  {\em JHEP} {\bf 0006} (2000) 002,
  [\href{http://xxx.lanl.gov/abs/hep-th/0004011}{{\tt hep-th/0004011}}].

\bibitem{Russo}
J.~G. Russo, {\it String spectrum of curved string backgrounds obtained by
  {T}--duality and shifts of polar angles},  {\em JHEP} {\bf 0509} (2005) 031,
  [\href{http://xxx.lanl.gov/abs/hep-th/0508125}{{\tt hep-th/0508125}}].

\bibitem{Spradlinetal}
M.~Spradlin, T.~Takayanagi, and A.~Volovich, {\it String theory in $\beta$
  deformed spacetimes},  {\em JHEP} {\bf 0511} (2005) 039,
  [\href{http://xxx.lanl.gov/abs/hep-th/0509036}{{\tt hep-th/0509036}}].

\bibitem{Kontsevich}
M.~Kontsevich, {\it Deformation quantization of {P}oisson manifolds {I}},  {\em
  Lett. Math. Phys.} {\bf 66} (2003) 157--216,
  [\href{http://xxx.lanl.gov/abs/q-alg/9709040}{{\tt q-alg/9709040}}].

\bibitem{BehrSykora}
W.~Behr and A.~Sykora, {\it Construction of {G}auge {T}heories on {C}urved
  {N}oncommutative {S}pacetime},  {\em Nucl. Phys. B} {\bf 698} (2004)
  473--502, [\href{http://xxx.lanl.gov/abs/hep-th/0309145}{{\tt
  hep-th/0309145}}].

\bibitem{Cerchiai}
B.-L. Cerchiai, {\it The {S}eiberg--{W}itten {M}ap for a {T}ime--dependent
  {B}ackground},  {\em JHEP} {\bf 0306} (2003) 056,
  [\href{http://xxx.lanl.gov/abs/hep-th/0304030}{{\tt hep-th/0304030}}].

\bibitem{Seibergetal}
N.~Seiberg, L.~Susskind, and N.~Toumbas, {\it Space/{T}ime
  {N}on--{C}ommutativity and {C}ausality},  {\em JHEP} {\bf 0006} (2000) 044,
  [\href{http://xxx.lanl.gov/abs/hep-th/0005015}{{\tt hep-th/0005015}}].

\bibitem{Gopakumaretal}
R.~Gopakumar, S.~M. J.~Maldacena, and A.~Strominger, {\it S--duality and
  noncommutative gauge theory},  {\em JHEP} {\bf 0006} (2000) 036,
  [\href{http://xxx.lanl.gov/abs/hep-th/0005048}{{\tt hep-th/0005048}}].

\bibitem{Gubseretal0009}
S.~S. Gubser, S.~Gukov, I.~R. Klebanov, M.~Rangamani, and E.~Witten, {\it The
  {H}agedorn transition in non--commutative open string theory},  {\em J. Math.
  Phys.} {\bf 42} (2001) 2749--2764,
  [\href{http://xxx.lanl.gov/abs/hep-th/0009140}{{\tt hep-th/0009140}}].

\bibitem{BarbonRabinovici}
J.~L.~F. Barbon and E.~Rabinovici, {\it On the {N}ature of the {H}agedorn
  {T}ransition in {NCOS} {S}ystems},  {\em JHEP} {\bf 0106} (2001) 029,
  [\href{http://xxx.lanl.gov/abs/hep-th/0104169}{{\tt hep-th/0104169}}].

\bibitem{RussoSheikhJabbari}
J.~G. Russo and M.~M. Sheikh-Jabbari, {\it On {N}oncommutative {O}pen {S}tring
  {T}heories},  {\em JHEP} {\bf 0007} (2000) 052,
  [\href{http://xxx.lanl.gov/abs/hep-th/0006202}{{\tt hep-th/0006202}}].

\bibitem{Niarchos00}
V.~Niarchos, {\it Density of {S}tates and {T}achyons in {O}pen and {C}losed
  {S}tring {T}heory},  {\em JHEP} {\bf 0106} (2001) 048,
  [\href{http://xxx.lanl.gov/abs/hep-th/0010154}{{\tt hep-th/0010154}}].

\bibitem{Sahakian01}
V.~Sahakian, {\it The large {M} limit of {N}on--{C}ommutative {O}pen {S}trings
  at strong coupling},  {\em Nucl. Phys. B} {\bf 621} (2002) 62--100,
  [\href{http://xxx.lanl.gov/abs/hep-th/0107180}{{\tt hep-th/0107180}}].

\bibitem{FriessGubser}
J.~J. Friess and S.~S. Gubser, {\it Instabilities of {D}--brane {B}ound
  {S}tates and {T}heir {R}elated {T}heories},  {\em JHEP} {\bf 0511} (2005)
  040, [\href{http://xxx.lanl.gov/abs/hep-th/0503193}{{\tt hep-th/0503193}}].

\bibitem{Aharonyetal00}
O.~Aharony, J.~Gomis, and T.~Mehen, {\it On {T}heories with {L}ight--{L}ike
  {N}oncommutativity},  {\em JHEP} {\bf 0009} (2000) 023,
  [\href{http://xxx.lanl.gov/abs/hep-th/0006236}{{\tt hep-th/0006236}}].

\bibitem{GomisMehen}
J.~Gomis and T.~Mehen, {\it Space--{T}ime {N}oncommutative {F}ield {T}heories
  {A}nd {U}nitarity},  {\em Nucl. Phys. B} {\bf 591} (2000) 259--276,
  [\href{http://xxx.lanl.gov/abs/hep-th/0005129}{{\tt hep-th/0005129}}].

\bibitem{HashimotoSethi}
A.~Hashimoto and S.~Sethi, {\it Holography and string dynamics in
  {T}ime--dependent backgrounds},  {\em Phys. Rev. Lett} {\bf 89} (2002)
  261601, [\href{http://xxx.lanl.gov/abs/hep-th/0208126}{{\tt
  hep-th/0208126}}].

\bibitem{AlishahihaParvizi}
M.~Alishahiha and S.~Parvizi, {\it Branes in {T}ime--{D}ependent {B}ackgrounds
  and {A}d{S}/{CFT} {C}orrespondence},  {\em JHEP} {\bf 0210} (2002) 047,
  [\href{http://xxx.lanl.gov/abs/hep-th/0208187}{{\tt hep-th/0208187}}].

\bibitem{Liuetal0204}
H.~Liu, G.~Moore, and N.~Seiberg, {\it Strings in a {T}ime--{D}ependent
  {O}rbifold},  {\em JHEP} {\bf 0606} (2006) 045,
  [\href{http://xxx.lanl.gov/abs/hep-th/0204168}{{\tt hep-th/0204168}}].

\bibitem{Liuetal0206}
H.~Liu, G.~Moore, and N.~Seiberg, {\it Strings in {T}ime--{D}ependent
  {O}rbifolds},  {\em JHEP} {\bf 0210} (2002) 031,
  [\href{http://xxx.lanl.gov/abs/hep-th/0206182}{{\tt hep-th/0206182}}].

\bibitem{Tai}
T.-S. Tai, {\it Winding {S}tring {D}ynamics in {T}ime--{D}ependent {B}eta
  {D}eformed {B}ackground},  \href{http://xxx.lanl.gov/abs/hep-th/0608136}{{\tt
  hep-th/0608136}}.

\bibitem{GregoryHarvey}
R.~Gregory and J.~Harvey, {\it Spacetime decay of cones at string coupling},
  {\em Class. Quant. Grav.} {\bf 20} (2003) L231--L238,
  [\href{http://xxx.lanl.gov/abs/hep-th/0306146}{{\tt hep-th/0306146}}].

\bibitem{Headrick}
M.~Headrick, {\it Decay of ${C}/{{Z}}_{n}$: exact supergravity solutions},
  {\em JHEP} {\bf 0403} (2004) 025,
  [\href{http://xxx.lanl.gov/abs/hep-th/0312213}{{\tt hep-th/0312213}}].

\bibitem{PilchWarner}
K.~Pilch and N.~P. Warner, {\it A {N}ew {S}upersymmetric {C}ompactification of
  {C}hiral {IIB} {S}upergravity},  {\em Phys. Lett. B} {\bf 487} (2000) 22--29,
  [\href{http://xxx.lanl.gov/abs/hep-th/0002192}{{\tt hep-th/0002192}}].

\bibitem{Halmagyietal}
N.~Halmagyi, K.~Pilch, C.~Romelsberger, and N.~P. Warner, {\it Holographic
  {D}uals of a {F}amily of ${\Ncal}=1$ {F}ixed {P}oints},  {\em JHEP} {\bf
  0608} (2006) 083, [\href{http://xxx.lanl.gov/abs/hep-th/0506206}{{\tt
  hep-th/0506206}}].

\bibitem{deBoeretal}
J.~de~Boer, P.~A. Grassi, and P.~van Nieuwenhuizen, {\it Non--commutative
  superspace from string theory},  {\em Phys. Lett. B} {\bf 574} (2003)
  98--104, [\href{http://xxx.lanl.gov/abs/hep-th/0302078}{{\tt
  hep-th/0302078}}].

\bibitem{OoguriVafa0302}
H.~Ooguri and C.~Vafa, {\it The {C}--{D}eformation of {G}luino and
  {N}on--planar {D}iagrams},  {\em Adv. Theor. Math. Phys.} {\bf 7} (2003)
  53--85, [\href{http://xxx.lanl.gov/abs/hep-th/0302109}{{\tt
  hep-th/0302109}}].

\bibitem{BerkovitsSeiberg}
N.~Berkovits and N.~Seiberg, {\it Superstrings in {G}raviphoton {B}ackground
  and ${\Ncal}=1/2+3/2$ {S}upersymmetry},  {\em JHEP} {\bf 0307} (2003) 010,
  [\href{http://xxx.lanl.gov/abs/hep-th/0306226}{{\tt hep-th/0306226}}].

\end{thebibliography}\endgroup
\bibliographystyle{JHEP}

\end{document}